\documentclass[12pt]{article}

\usepackage{amssymb}
\usepackage{amsmath}
\usepackage{amscd}
\usepackage{latexsym}
\usepackage{graphicx}
\usepackage{color}

\usepackage{cite}

\newcommand{\be}{\begin{equation}}
\newcommand{\ee}{\end{equation}}
\newcommand{\Dlt}{\Delta}
\newcommand{\dlt}{\delta}

\newcommand{\br}{{\bf r}}

\newcommand{\bfe}{{\bf e}}
\newcommand{\bn}{{\bf n}}
\newcommand{\bB}{{\bf B}}
\newcommand{\bS}{{\bf S}}
\newcommand{\bt}{\beta}

\newcommand{\al}{\alpha}

\newcommand{\gm}{\gamma}
\newcommand{\om}{\omega}

\newcommand{\rgl}{\rangle}
\newcommand{\lgl}{\langle}

\begin{document}

\begin{center}

{\Large{\bf Spintronics with magnetic nanomolecules and graphene flakes} \\ [5mm]

V.I. Yukalov$^{1,*}$, V.K. Henner$^{2,3}$, T.S. Belozerova$^2$, and \\ [2mm]
E.P. Yukalova$^4$} \\ [3mm]

{\it

$^1$Bogolubov Laboratory of Theoretical Physics, \\
Joint Institute for Nuclear Research, Dubna 141980, Russia \\ [3mm]

$^2$Department of Physics, Perm State University, Perm 614990, Russia \\ [3mm]

$^3$Department of Physics, University of Louisville, Louisville,
Kentucky 40292, USA \\ [3mm]

$^4$Laboratory of Information Technologies, \\
Joint Institute for Nuclear Research, Dubna 141980, Russia }

\end{center}

\vskip 2cm

\begin{abstract}

We show how the magnetization of nano-objects can be efficiently regulated.
Several types of nanosystems are considered: magnetics nanomolecules, 
magnetic nanoclusters, polarized nanomolecules, and magnetic graphene. 
These nano-objects and the structures composed of them enjoy many common 
properties, with the main difference being in the type of particle 
interactions. The possibility of governing spin dynamics is important for 
spintronics. 
\end{abstract}

\vskip 1.5cm
{\parindent=0pt
{\bf Keywords}: Magnetic nanomolecules; Magnetic nanoclusters; Polarized
nanomolecules; Magnetic graphene; Spintronics

\vskip 2.5cm

{\bf Corresponding author}: \\
V.I. Yukalov \\
Bogolubov Laboratory of Theoretical Physics \\
Joint Institute for Nuclear Research, Dubna, Russia \\
{\bf e-mail}: yukalov@theor.jinr.ru

}

\newpage

\section{Introduction}

Several types of nano-objects possess magnetic moments. For instance,
these are magnetic nanomolecules having a finite spin polarization in 
their ground state \cite{Kahn_1,Barbara_2,Yukalov_3,Yukalov_4}. The 
typical size of a nanomolecule is $1$ nm. In the ground state, magnetic 
nanomolecules have spins $S \sim 1/2 - 27/2$ that are frozen below the 
blocking temperature $T_B \sim 1 - 10$ K. Magnetic nanoclusters, formed 
by magnetic atoms, are very similar to magnetic molecules 
\cite{Yukalov_3,Yukalov_4,Kodama_5,Hadjipanays_6,Wernsdorfer_7,Ferre_8,
Bedanta_9,Berry_10,Beveridge_11}. The typical sizes of such nanoclusters 
are defined by the coherence radius $R_{coh} \sim 1 -10$ nm, sometimes up 
to $R_{coh} \sim 100$ nm. Below the blocking temperature 
$T_B \sim 10 - 100$ K, the magnetization is frozen. Magnetic nanoclusters
can contain a large number of atoms $N \sim 100 - 10^5$, which defines 
large values of their spins $S \sim 100 - 10^5$. Note that nanoclusters 
of the size smaller than the coherence radius behave as a single magnetic 
molecule. But if the size is larger than the coherence radius, a nanocluster 
separates into several magnetic domains with opposite magnetization making 
the total spin zero. There also exists a class of nanomolecules containing 
a large number of hydrogen atoms, whose protons can be polarized. Examples 
are propanediol, butanol, ammonia, and many biological molecules. Systems 
of such molecules, although having no polarization in their ground state, 
but can be polarized and at low temperatures can sustain polarization for 
very long time \cite{Yukalov_3,Yukalov_4,Chen_12,Krishnan_13}. Recently, 
there has appeared a novel class of nanomaterials that can possess magnetic 
moments. These are graphene flakes, graphene ribbons, and carbon nanotubes 
with defects \cite{Yaziev_14,Katsnelson_15,Enoki_16}.     

The possibility of efficiently regulating spin dynamics and governing spin 
directions is important for quantum information processing and different 
applications in spintronics. It is admissible to influence spin motion by 
transverse external fields or laser beams. In the present paper, we consider 
a principally different method based on the existence of a feedback field 
caused by the coupling of a magnetic sample with a resonant electric circuit.   
This way serves as the easiest mechanism of regulating spin dynamics and
provides the fastest method for realizing spin reversal.

\section{Resonant electric circuit}

The sample, containing magnetic particles, with magnetic moment $\mu_0$, 
is inserted into a magnetic coil of $n$ turns, length $l$, and volume $V_c$. 
The coil is a part of an electric circuit, with capacitance $C$, induction $L$, 
and resistance $R$. The coil axis is along the axis $x$. Electric current 
in the circuit is defined by the Kirchhoff equation 
\be
\label{1}
 L \; \frac{dj}{dt} + R j + \frac{1}{C} \int_0^t j(t)\; dt = - \; 
\frac{d\Phi}{dt} \; ,
\ee
where the current $j$ is generated by the magnetic flux
\be
\label{2}
 \Phi = \frac{4\pi n}{c l} \; M_x \;  ,
\ee
caused by the $x$-component of the magnetization
\be
\label{3}
 M_x = \mu_0 \sum_{j=1}^N \; \lgl S_j^x \rgl \;  .
\ee
The electric current in the coil creates magnetic field directed along 
the axis $x$ and having the magnitude
\be
\label{4}
H = \frac{4\pi n}{c l} \; j
\ee
characterized by the feedback equation
\be
\label{5}
 \frac{dH}{dt} + 2\gm H + \om^2 \int_0^t H(t')\; dt = - 4\pi \; 
\frac{d m_x}{dt} \;  ,
\ee
where $\omega$ is the circuit natural frequency and $\gamma$ is the 
circuit damping,
\be
\label{6}
 \om \equiv \frac{1}{\sqrt{L C}} \; , \qquad 
\gm \equiv \frac{R}{2L} \;  .
\ee
The electromotive force in the right-hand side is produced by the 
moving magnetization density 
\be
\label{7}
 m_x \equiv \frac{M_x}{V_c} = \frac{\mu_0}{V_c} 
\sum_{j=1}^N \; \lgl S_j^x \rgl \;  ,
\ee
with $V_c$ being the coil volume. 

The sample is subject to an external magnetic field $B_0$ directed along
the axis $z$. Thus, the total magnetic field, acting on the sample, is
\be
\label{8}
 \bB = H \bfe_x + B_0 \bfe_z \;  .
\ee
The external magnetic field defines the Zeeman frequency
\be
\label{9}
 \om_0 \equiv \frac{|\mu_0 B_0|}{\hbar} \;  .
\ee

The electric circuit is called a resonator, since its natural frequency 
is tuned close to the Zeeman frequency, so that the resonance condition
\be
\label{10}
\frac{|\Dlt|}{\om} \; \ll \; 1 \qquad ( \Dlt \equiv \om - \om_0 )
\ee
be valid.

\section{Magnetic nanomolecules and nanoclusters}

A system of $N$ magnetic nanomolecules or magnetic nanoclusters is described 
by the Hamiltonian
\be
\label{11}
 \hat H = \sum_i \hat H_i + 
\frac{1}{2} \sum_{i\neq j} \hat H_{ij} \;  .
\ee
Here the single-molecule Hamiltonian is
\be
\label{12}
 \hat H_i = - \mu_0 \bB \cdot \bS_i - D \left ( S_i^z \right )^2 +
 D_2 \left ( S_i^x \right )^2 + D_4 \left [ 
\left ( S_i^x \right )^2 \left ( S_i^y \right )^2 + 
\left ( S_i^y \right )^2 \left ( S_i^z \right )^2 +
\left ( S_i^z \right )^2 \left ( S_i^x \right )^2 \right ] \; ,
\ee
in which $S_i^\alpha$ is a spin-operator component of an $i$-th molecule and 
$D$, $D_2$, and $D_4$ are anisotropy parameters. The molecular interactions
are of dipolar origin, with the interaction Hamiltonian
\be
\label{13}
 \hat H_{ij}  = \sum_{\al\bt} D_{ij}^{\al\bt} S_i^\al S_j^\bt ,
\ee
in which  
$$
D_{ij}^{\al\bt} = \frac{\mu_0^2}{r_{ij}^3} \; \left ( \dlt_{\al\bt} -
3 n_{ij}^\al n_{ij}^\bt \right )
$$
is the dipolar tensor and
$$
 r_{ij} \equiv | \br_{ij} | \; , \qquad 
\bn_{ij} \equiv \frac{\br_{ij}}{r_{ij}} \; , \qquad 
\br_{ij} \equiv \br_i - \br_j \; .
$$

Suppose that the system has been polarized and placed in an external magnetic 
field with the direction opposite to the equilibrium state. This means that
the sample is prepared in a strongly nonequilibrium initial state. When no
pushing transverse field is imposed, the spin motion is triggered by spin
waves \cite{Yukalov_4,Yukalov_17,Yukalov_18}. Note that spin waves can be well
defined for arbitrary nonequilibrium states \cite{Birman_19}.   

Sometimes, one discusses the possibility of two other triggering mechanisms,
thermal noise in the coil and photon exchange between moving spins. First, 
let us evaluate the influence on the spin motion of the coil thermal noise. 
The characteristic {\it thermal time}, required for the thermal noise to start 
moving spins \cite{Yukalov_4} is
\be
\label{14}  
 t_T = \frac{4\hbar\gm V_c}{\mu_0^2 \om} \; 
\tanh \left ( \frac{\om}{2\om_T} \right ) \;  ,
\ee
where 
$$
\om_T \equiv \frac{k_B T}{\hbar}
$$
is the thermal frequency. When the latter is larger than the resonator 
natural frequency, then the thermal time is
\be
\label{15}
 t_T \simeq \frac{2\hbar\gm V_c}{\mu_0^2 \om_T} \qquad
\left ( \frac{\om}{\om_T} \ll 1 \right ) \;  .
\ee
The coupling of the magnet with the coil induces the {\it coupling attenuation}
$$
 \gm_c = \pi \mu_0^2 \; \frac{ N S}{\hbar V_c} \; ,
$$
due to the coil feedback field. Therefore the influence of the coil on spins 
arises after the {\it coupling time} that can be evaluated as
\be
\label{16}
 t_c \equiv \frac{1}{\gm_c} = \frac{\hbar V_c}{\pi \mu_0^2 N S} \;  .
\ee
The ratio of the thermal to coupling times becomes
\be
\label{17}
 \frac{t_T}{t_c} \approx \frac{2\pi \gm}{\om_T} \; N S \;  .
\ee
For magnetic nanomolecules and nanoclusters, the thermal time is much longer
than the coupling time, hence thermal noise does not play role for initiating
spin motion. For instance, at $T = 1$ K, we have $\omega_T \sim 10^{12}$ s$^{-1}$.
Then the thermal noise could be important only either for a small number of
molecules or for small spin values, such that $2 \pi \gamma N S \ll 10^{12}$.
Thus, if $\gamma \sim 10^{10}$ s$^{-1}$, then it should be $N S  \ll 20$. 
Although for small number of low-spin particles, the thermal noise could play
role \cite{Kiselev_20}, but for magnetic nanomolecules with $S \sim 10$ and 
moreover for magnetic nanoclusters with $S \sim 100 - 10^5$, thermal noise 
is not able to influence spin motion even for a single particle.   

The other mechanism that sometimes is assumed to be present in collectivizing
the rotational motion of spins in a solid-state system is the Dicke effect 
caused by the photon exchange between moving spins, similarly to the 
developing coherence and superradiance in an ensemble of radiating atoms 
\cite{Allen_21} or ions \cite{Bourhill_22}. Moving spins really represent 
dipole emitters of photons \cite{Yukalov_23,Yukalov_24}, with the natural width 
$$
 \gm_0 =  \frac{2\mu_0^2 S k^3}{3\hbar} \qquad 
\left ( k \equiv \frac{\om}{c} \right ) \; .
$$
The moving spins could induce mutual correlations in the spin motion after
the typical radiation time
\be
\label{18}   
 t_{rad} \equiv \frac{1}{\gm_0} = \frac{3\hbar}{2\mu_0^2 S k^3} \; .
\ee
However, for magnetic nanomolecules or nanoclusters, in the magnetic field
$B_0 = 1 T = 10^4 G$, we have $\omega_0 \sim 10^{11}$ s$^{-1}$. Then this 
radiation time is extremely long, being of order 
$t_{rad} \sim 10^8$ s $\sim 10$ years. On the other hand, the dephasing time 
due to dipolar forces is
\be
\label{19}
 T_2 \equiv \frac{1}{\gm_2} = \frac{\hbar}{\rho\mu_0^2 S} \;  ,
\ee
where $\rho \sim a^{-3}$ is the density of spins, with $a$ being the mean
interspin distance. For nanoclusters, with $\rho \sim 10^{20}$ cm$^{-3}$
and $S \sim 10^3$, the dephasing time is $T_2 \sim 10^{-10}$ s. The radiation 
time, related to photons, is incomparably longer than the dephasing time,
\be
\label{20}
  \frac{t_{rad}}{T_2} = \frac{3\rho}{2k} \gg 1 \; ,
\ee
being of order $10^{18}$ and making the correlations due to photon exchange 
absolutely irrelevant for spin systems \cite{Yukalov_24,Yukalov_25}.

The motion of spins in magnets is triggered by spin waves and mutual 
correlations can be induced only by the resonator feedback field 
\cite{Yukalov_3,Yukalov_4,Yukalov_24,Yukalov_25}. The amplifying role of 
the resonator is called the Purcell effect \cite{Purcell_26}. Such an  
amplification is known in the case of magnetic resonance 
\cite{Chen_12,Krishnan_13,Davis_27}

The anisotropy parameters for nanoclusters, such as Co, Fe, and Ni, as 
compared to $\gamma_2 \equiv 1/T_2$, are
$$
 \frac{D}{\hbar\gm_2} \sim 10^{-3} \; , \qquad 
\frac{D_2}{\hbar\gm_2} \sim 10^{-3} \; , \qquad
\frac{D_4}{\hbar\gm_2} \sim 10^{-10} \;  .
$$

We have accomplished numerical investigation of spin dynamics for a system 
of many spins and for separate spins. For spin systems, we consider the 
average spin polarizations, such as the $z$-polarization
$$
 s \equiv \frac{1}{NS} \sum_{j=1}^N \lgl S_j^z \rgl \;  .
$$ 
The spin dynamics of a single nanocluster ($N = 1$) as a function of 
dimensionless time, measured in units of $1/\gamma_2$, is represented in 
Fig. 1, where the notation $h \neq 0$ implies the presence of a resonator 
producing the feedback field $h \equiv \mu_0 H/ \hbar \gamma_c$, while the 
notation $h = 0$ means the absence of the resonator. As it should be, 
there is no spin reversal without a resonator. While in the presence of 
the resonator there occurs a very fast spin reversal during the reversal 
time $t_{rev} \sim 10^{-10}$ s. Since, for a single nanocluster, there are
no spin waves, the initial spin motion is triggered by a weak transverse
magnetic field. 

For many nanoclusters or nanomolecules, the reversal time becomes even 
shorter, reaching $t_{rev} \sim 10^{-12}$ s. One of the main differences 
between nanomolecules and nanoclusters is in the following. Nanoclusters
cannot be prepared being identical in their sizes and spin values. While
nanomolecules of the same chemical structure are identical with each other
and can form perfect crystalline lattices. However, the nonuniformity of
nanoclusters does not preclude the possibility of their spin correlations
by the resonator feedback field \cite{Kharebov_28}.

\section{Magnetic graphene flakes}

A novel interesting nanomaterial, enjoying magnetic properties, is graphene 
with defects, which can be in the form of graphene flakes, graphene ribbons, 
and carbon nanotubes with defects \cite{Yaziev_14,Katsnelson_15,Enoki_16}.
Spins are located at defects that can be of various nature. The principal
difference of such a magnetic graphene from the case of magnetic 
nanomolecules and nanoclusters is in the nature of spin interactions. In
magnetic graphene, spins interact with each other by {\it exchange 
interactions}, so that the Hamiltonian reads as
\be
\label{21}
 \hat H = - \mu_0 \sum_{j=1}^N \bB \cdot \bS_j + 
\frac{1}{2} \sum_{i\neq j}^N \hat H_{ij} \;  ,
\ee
where the exchange interactions have the anisotropic Heisenberg form
\be
\label{22}
 \hat H_{ij} = - J_{ij} \left ( S_i^x S_j^x + S_i^y S_j^y \right ) -
I_{ij} S_i^z S_j^z \; .
\ee       

Since there are a number of ways for creating magnetic defects in graphene,
the interaction parameters can be varied 
\cite{Yaziev_14,Katsnelson_15,Enoki_16}. The governing of spin dynamics is
realized similar to the case of nanomolecules and nanoclusters. The physics 
of the processes is also similar. We study a graphene flake prepared in a 
strongly nonequilibrium initial state, with the spin polarization opposite 
to the equilibrium direction in an external magnetic field. Magnetic defects 
are located on a zigzag edge of graphene. Up to $100$ defects are considered.
The spin motion is triggered by spin waves and collectivized by the resonator 
feedback field. Because of the developed coherence in spin motion, the spin 
reversal happens in short time of order $t_{rev}\sim 10^{-11}$ s.  

We investigate the spin dynamics of defects in graphene for different 
parameters of the anisotropy
\be
\label{23}
 \al \equiv \frac{I_{ij}}{J_{ij}} \;  ,
\ee
where the nearest-neighbor interactions are taken, for different coupling 
parameters
\be
\label{24}
 \bt \equiv \left | \; \frac{\hbar\gm_c}{\pi\om_0} \; \right | =
 \left | \; \frac{\mu_0 NS}{B_0 V_c} \; \right | \; ,   
\ee
under the resonator quality factor
\be
\label{25}
 Q \equiv \frac{\om}{2\gm} \;  .
\ee

Figure 2 demonstrates the behavior of the average spin polarizations 
$e_x$ and $e_z$, defined as
$$
 e_\al \equiv \frac{1}{NS} \sum_{j=1}^N \lgl S_j^\al \rgl \qquad
( \al = x,\; z ) \;  ,
$$
for different magnitudes of the magnetic anisotropy. As is seen, the 
larger the magnetic anisotropy, the longer the reversal time. Figure 3
shows the spin polarizations for the same parameters, as in Fig. 2, except
for the interaction range that is taken to connect not merely the nearest 
neighbors, but $3$ neighbor shells. The increasing interaction radius 
influences the most the sample with a strong magnetic anisotropy.

\section{Conclusion}

\begin{sloppypar}
We have studied spin dynamics for several types of nanomaterials: magnetic
nanomolecules, magnetic nanoclusters, and magnetic graphene flakes. In the 
latter, localized magnetic moments arise because of incorporated defects.
The difference between magnetic nanomolecules or nanoclusters and magnetic 
graphene is in their magnetic interactions. Magnetic nanomolecules and 
nanoclusters interact with each other through dipolar forces, while magnetic
defects in graphene exhibit exchange interactions.
\end{sloppypar}

For all types of magnetic samples, we consider a similar setup. The system
is initially prepared in a strongly nonequilibrium state, with a magnetization
opposite to the equilibrium one, in the presence of an external magnetic field.
The sample is inserted into a magnetic coil of a resonant electric circuit.
When there are no transverse external fields, the spin motion is triggered by
spin waves. The collective dynamics develops because of the resonator feedback 
field. The influence of a resonator on spin motion is what is termed the 
Purcell effect. Coherent spin motion arises due to the Purcell effect. While 
the Dicke effect of interactions through the photon exchange plays no role 
for spin systems. The resonator thermal noise can influence spin dynamics only
for small number of low-spin assemblies. But for large spin systems, as well as 
for nanomolecules or nanoclusters with high spins, the resonator thermal noise 
also is not important.        

By varying the parameters of the resonant electric circuit and external 
magnetic fields, it is possible to realize efficient manipulation of spin 
dynamics, which can be used in various applications of spintronics, such 
as magnetic data storage and data processing \cite{Dimian_29}. The considered 
mechanism for the fast regulation of spin dynamics can also be employed for 
biological molecules and magnetotactic bacteria \cite{Wang_30}.

\newpage

\begin{center}

{\Large {\bf Figure Captions }}

\end{center}

\vskip 3cm
{\bf Figure 1}. Spin polarization $s = s(t)$ of a single nanocluster as 
a function of the dimensionless time, measured in units of $1/\gamma_c$, 
for the parameters $\omega/\gamma_c = \omega_0/\gamma_c = 10$, 
$(2S - 1)D/(\hbar \gamma_c) = (2S - 1)D_2/(\hbar \gamma_c) = 1$,
and small $D_4/D \sim 10^{-7}$, with $\gamma/\gamma_c = 1$. The solution
in the presence of resonator (solid line) is compared with that in the 
absence of resonator (dashed line). 

\vskip 1cm
{\bf Figure 2}. Transverse polarization $e_x$ (dotted line) and longitudinal 
polarization $e_z$ (solid line) for magnetic graphene with $100$ defects on 
a zigzag edge as a function of dimensionless time, measured in units of 
$1/\omega$, for the coupling parameter $\beta = 0.01$, the resonator quality
factor $Q = 10$, and different anisotropy parameters: (a) $\alpha = 1$; 
(b) $\alpha = 1.2$; (c) $\alpha = 1.4$. Only the nearest-neighbor interactions
are assumed. 

\vskip 1cm
{\bf Figure 3}. The same as in Fig. 2, but for interactions ranging over 
three neighbor shells.

\newpage

\begin{figure}[ht!]
\vspace{9pt}
\centerline{
\includegraphics[width=8cm]{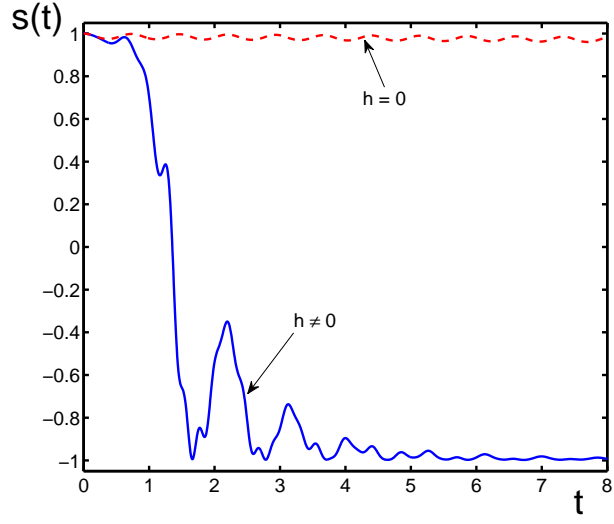} }
\caption{Spin polarization $s = s(t)$ of a single nanocluster as a 
function of the dimensionless time, measured in units of $1/\gamma_c$, 
for the parameters $\omega/\gamma_c = \omega_0/\gamma_c = 10$, 
$(2S - 1)D/(\hbar \gamma_c) = (2S - 1)D_2/(\hbar \gamma_c) = 1$,
and small $D_4/D \sim 10^{-7}$, with $\gamma/\gamma_c = 1$. The solution
in the presence of resonator (solid line) is compared with that in the 
absence of resonator (dashed line).
}
\label{fig:Fig.1}
\end{figure}

\newpage

\begin{figure}[ht!]
\vspace{9pt}
\centerline{
\includegraphics[width=12cm]{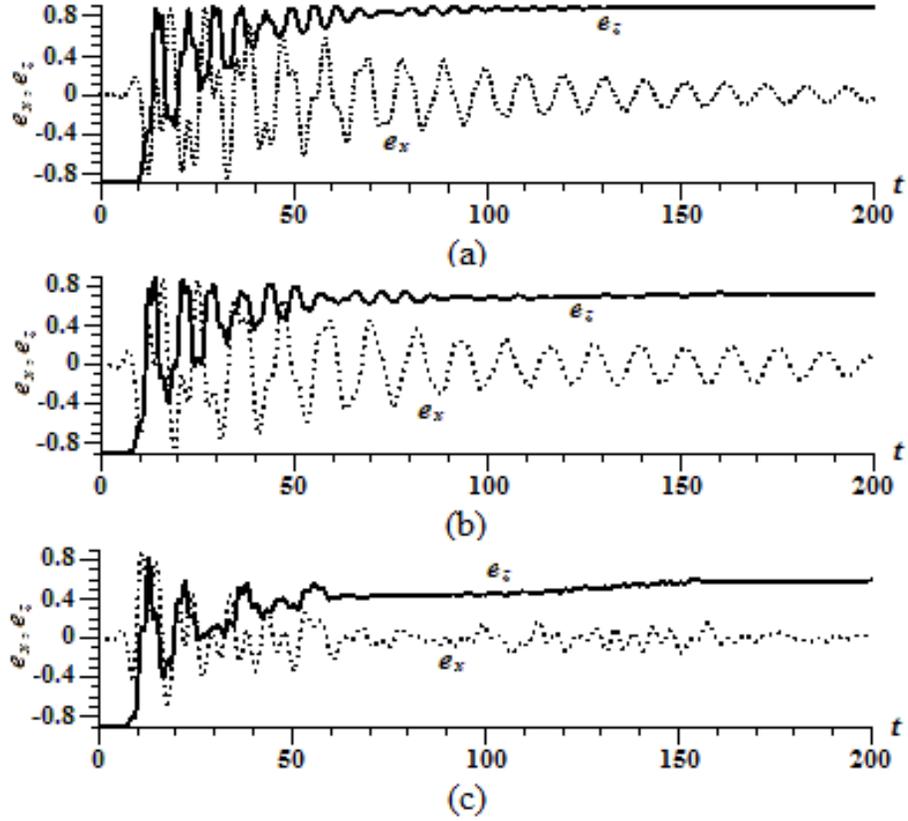} }
\caption{Transverse polarization $e_x$ (dotted line) and longitudinal 
polarization $e_z$ (solid line) for magnetic graphene with $100$ defects on 
a zigzag edge as a function of dimensionless time, measured in units of 
$1/\omega$, for the coupling parameter $\beta = 0.01$, the resonator quality
factor $Q = 10$, and different anisotropy parameters: (a) $\alpha = 1$; 
(b) $\alpha = 1.2$; (c) $\alpha = 1.4$. Only the nearest-neighbor interactions
are assumed. 
}
\label{fig:Fig.2}
\end{figure}

\newpage

\begin{figure}[ht!]
\vspace{9pt}
\centerline{
\includegraphics[width=12cm]{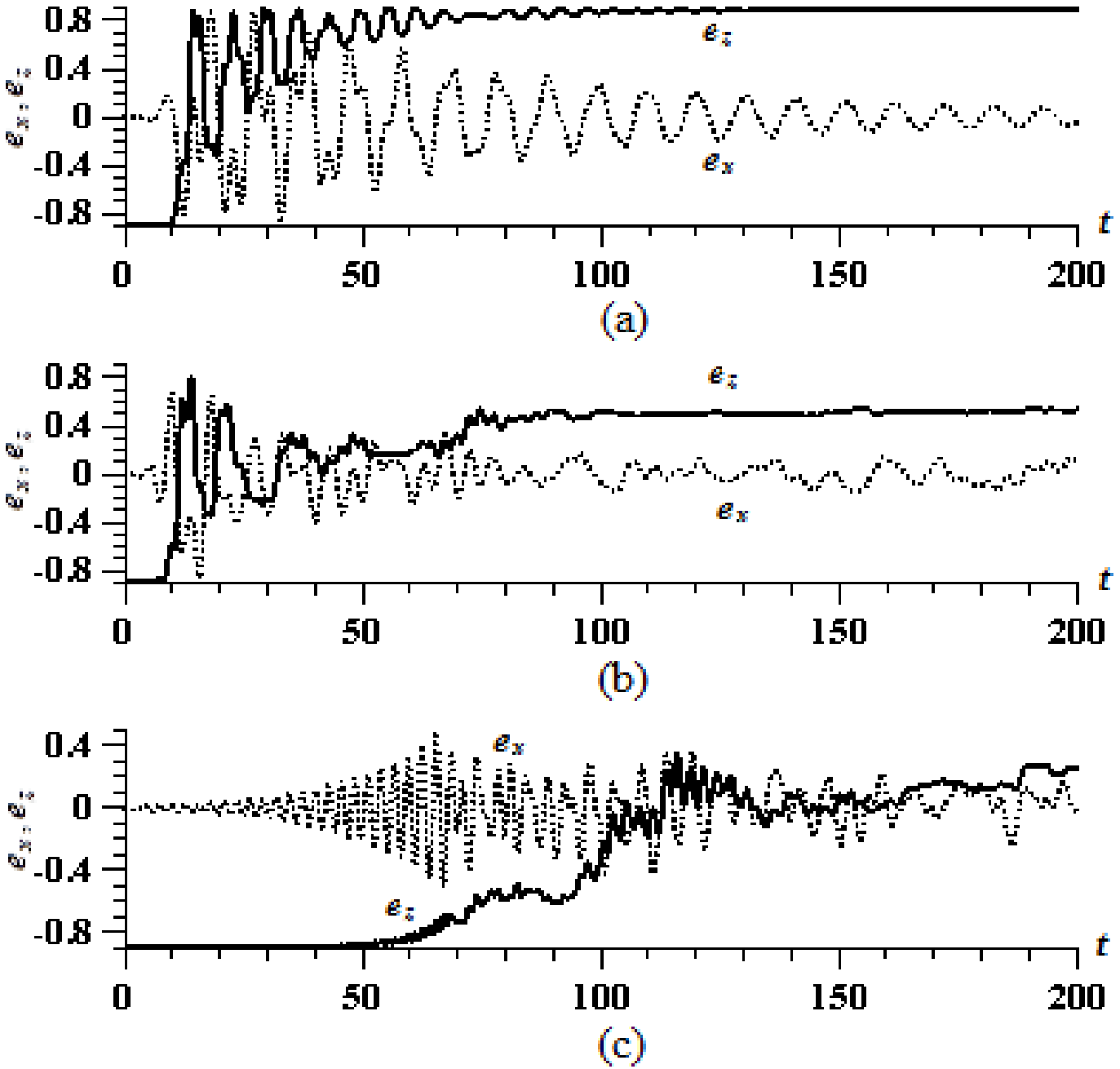} }
\caption{The same as in Fig. 2, but for interactions ranging over 
three neighbor shells.
}
\label{fig:Fig.3}
\end{figure}


\newpage


\begin{thebibliography}{99}

\bibitem{Kahn_1}
Kahn, O.: 
Molecular Magnetism. VCH, New York (1995)

\bibitem{Barbara_2}
Barbara, B., Thomas, L., Lionti, F., Chioresku, I., Sulpice, A.:
Macroscopic quantum tunneling in molecular magnets.
J. Magn. Magn. Mater. {\bf 200}, 167--182  (1999)

\bibitem{Yukalov_3}
Yukalov, V.I.:
Superradiant operation of spin masers.
Laser Phys. {\bf 12}, 1089--1103 (2002)

\bibitem{Yukalov_4}
Yukalov, V.I., Yukalova, E.P.:
Coherent nuclear radiation.
Phys. Part. Nucl. {\bf 35}, 348--382 (2004) 

\bibitem{Kodama_5}
Kodama, R.H.:
Magnetic nanoparticles.
J. Magn. Magn. Mater. {\bf 200}, 359--372 (1999)

\bibitem{Hadjipanays_6}
Hadjipanays, G.C.:
Nanophase hard magnets.
J. Magn. Magn. Mater. {\bf 200}, 373-- (1999)

\bibitem{Wernsdorfer_7}
Wernsdorfer, W.:
Classical and quantum magnetization reversal studied in nanometer-sized
particles and clusters.
Adv. Chem. Phys. {\bf 118}, 99--190 (2001) 

\bibitem{Ferre_8}
Ferre, J.:
Dynamics of magnetization reversal: from continuous to patterned
ferromagnetic films.
Adv. Chem. Phys. {\bf 83}, 127--185 (2002) 

\bibitem{Bedanta_9}
Bedanta, S., Kleemann, W.:
Supermagnetism.
J. Phys. D {\bf 42}, 013001 (2009)

\bibitem{Berry_10}
Berry, C.C.:
Progress in functionalization of magnetic nanoparticles for application 
in biomedicine.
J. Phys. D {\bf 42}, 224003 (2009)

\bibitem{Beveridge_11}
Beveridge, J.S., Stephens, J.R., Willimas, M.E.:
The use of magnetic nanoparticles in analytical chemistry.
Annu. Rev. Anal. Chem. {\bf 4}, 251--273 (2011)

\bibitem{Chen_12}
Chen, H.Y., Lee, Y., Bowen, S., Hilty, C.:
Spontaneous emission of NMR signals in hyperpolarized proton spin systems.
J. Magn. Res. {\bf 208}, 204--209 (2011) 

\bibitem{Krishnan_13}
Krishnan, V.V., Murali, N.:
Radiation damping in modern NMR experiments: progress and chalenges.
Prog. Nucl. Magn. Res. Spectrosc. {\bf 68}, 41--57 (2013).

\bibitem{Yaziev_14}
Yaziev, O.V.:
Emergence of magnetism in graphene materials and nanostructures.
Rep. Prog. Phys. {\bf 73}, 056501 (2010) 

\bibitem{Katsnelson_15}
Katsnelson, M.I.:
Graphene: Carbon in Two Dimensions. 
Cambridge University, Cambridge (2012)

\bibitem{Enoki_16}
Enoki, T., Ando, T.:
Physics and Chemistry of Graphene. 
Pan Stanford, Singapore (2013)

\bibitem{Yukalov_17}
Yukalov, V.I.:
Origin of pure spin superradiance.
Phys. Rev. Lett. {\bf 75}, 3000--3003 (1995)

\bibitem{Yukalov_18}
Yukalov, V.I.:
Nonlinear spin dynamics in nuclear magnets.
Phys. Rev. B {\bf 53}, 9232--9250 (1996)

\bibitem{Birman_19}
Birman, J.L., Nazmitdinov, R.G., Yukalov, V.I.:
Effects of symmetry breaking in finite quantum systems.
Phys. Rep. {\bf 526}, 1--91 (2013)

\bibitem{Kiselev_20}
Kiselev, Y.F., Shumovsky, A.S., Yukalov, V.I.:
Thermal-noise induced radio frequency superradiance in resonator.
Mod. Phys. Lett. B {\bf 3}, 1149--1156 (1989)

\bibitem{Allen_21}
Allen, L., Eberly, J.H.:
Optical Resonance and Two-Level Atoms.
Wiley, New York (1975)

\bibitem{Bourhill_22}
Bourhill, J., Goryachev, M., Farr, W.G., Tobar, M.E.:
Superradiant behavior of $Cr^{3+}$ ions in ruby revealed by whispering 
gallery modes.
arXiv:1504.07733 (2015)

\bibitem{Yukalov_23}
Yukalov, V.I.:
Coherent dynamics of radiating atomic systems in pseudospin representation.
Laser Phys. {\bf 24}, 094015 (2014)

\bibitem{Yukalov_24}
Yukalov, V.I., Yukalova, E.P.:
Absence of spin superradiance in resonatorless magnets.
Laser Phys. Lett. {\bf 2}, 302--308 (2005)

\bibitem{Yukalov_25}
Yukalov, V.I.:
Spin superradiance versus atomic superradiance.
Laser Phys. Lett. {\bf 2}, 356--361 (2005)

\bibitem{Purcell_26}
Purcell, E.M.:
Spontaneous emission probabilities at radio frequencies.
Phys. Rev. {\bf 69}, 681 (1946)

\bibitem{Davis_27}
Davis, C.L., Kaganov, I.V., Henner, V.K.:
Superradiation in magnetic resonance.
Phys. Rev. B {\bf 62}, 12328--12337 (2000) 

\bibitem{Kharebov_28}
Kharebov, P.V., Henner, V.K., Yukalov, V.I.:
Optimal conditions for magnetization reversal of nanocluster assemblies
with random properties. 
J. Appl. Phys. {\bf 113}, 043902 (2013)

\bibitem{Dimian_29}
Dimian, M.:
Nonlinear spin dynamics and ultra-fast precessional switching.
Ph.D. Thesis, University of Maryland (2005) 

\bibitem{Wang_30}
Wang, H., Yu, Y., Sun, Y., Chen, Q.:
Magnetic nanochains.
Nano {\bf 6}, 1--17 (2011)   



\end{thebibliography}
\end{document}